\documentclass{elsart}

\usepackage{amssymb}
\usepackage{bm}%
\usepackage[colorlinks=true,linkcolor=red]{hyperref}
\usepackage{amsmath}
\usepackage{amsfonts}
\usepackage{amssymb}

\usepackage{graphicx}
\usepackage[sort&compress]{natbib}

\newtheorem{Task}{Task}
\newtheorem{Remark}{Remark}
\def\Tr{\mathrm{Tr}}
\newcounter{myfootertablecounter}

\newcommand\myfootnotemark{%
\addtocounter{footnote}{1}%
\footnotemark[\thefootnote]%
}%

\newcommand\myfootnotetext[1]{%
\addtocounter{myfootertablecounter}{1}
\footnotetext[\value{myfootertablecounter}]{#1} }

\newcommand\myfootnote[1]{%
\addtocounter{myfootertablecounter}{1} \footnote{#1}
}%

\begin{document}

\begin{frontmatter}


 \title{On invariant $2\times 2$ $\beta$-ensembles of random matrices.}
 \author{Pierpaolo Vivo\corauthref{cor1}}
 \ead{pierpaolo.vivo@brunel.ac.uk}
 \corauth[cor1]{Corresponding author.}
 \address{School of Information Systems, Computing \&
Mathematics\\Brunel University, Uxbridge, Middlesex, UB8
3PH\\United Kingdom}
 \author{Satya N. Majumdar}
 \ead{satya.majumdar@u-psud.fr }
 \address{Laboratoire de Physique Th\'{e}orique et Mod\`{e}les
Statistiques (UMR 8626 du CNRS)\\ Universit\'{e} Paris-Sud,
B\^{a}timent 100, 91405 Orsay Cedex\\ France}

\title{}


\author{}

\address{}

\begin{abstract}
We introduce and solve exactly a family of invariant $2\times 2$
random matrices, depending on one parameter $\eta$, and we show
that rotational invariance and real Dyson index $\beta$ are not
incompatible properties. The probability density for the entries
contains a weight function and a multiple trace-trace interaction
term, which corresponds to the representation of the
Vandermonde-squared coupling on the basis of power sums. As a
result, the effective Dyson index $\beta_{\mathrm{eff}}$ of the
ensemble can take any real value in an interval. Two weight
functions (Gaussian and non-Gaussian) are explored in detail and
the connections with $\beta$-ensembles of Dumitriu-Edelman and the
so-called Poisson-Wigner crossover for the level spacing are
respectively highlighted. A curious spectral twinning between
ensembles of different symmetry classes is unveiled: as a
consequence, the identification between symmetry group
(orthogonal, unitary or symplectic) and the exponent of the
Vandermonde ($\beta=1,2,4$) is shown to be potentially deceptive.
The proposed technical tool more generically allows for designing
actual matrix models which i) are rotationally invariant; ii) have
a real Dyson index $\beta_{\mathrm{eff}}$; iii) have a
pre-assigned confining potential or alternatively level-spacing
profile. The analytical results have been checked through
numerical simulations with an excellent agreement. Eventually, we
discuss possible generalizations and further directions of
research.
\end{abstract}

\end{frontmatter}

\section{Introduction.}

Ensembles of matrices with random elements have been widely
studied since the pioneering works of Wigner \cite{Wigner} and
Dyson \cite{Dys:new} on the 'threefold way'. A first, gross
classification of random matrix (RM) models can take into account
i) whether the size $N$ of the matrices in the ensemble is finite
or the limit $N\rightarrow\infty$ is taken and ii) whether the
probability distribution of the entries remains invariant after a
rotation in the matrix space.

The requirement of rotational invariance implies that the joint
probability density (jpd) of the eigenvalues can be written as:
\begin{equation}\label{jpdGauss}
  P(\lambda_1,\ldots,\lambda_N)\propto e^{-\frac{1}{2}\sum_{i=1}^N
  V(\lambda_i)}\prod_{j<k}|\lambda_j-\lambda_k|^\beta
\end{equation}
where $V(x)$ is a confining potential ($x^2$ for Gaussian
ensembles) and the interaction term between eigenvalues is the
well-known Vandermonde determinant raised to the power $\beta$.
The Dyson index $\beta$ can classically take \emph{only} the
values $1,2,4$ according to the number of variables needed to
specify a single entry ($1$ for real, $2$ for complex and $4$ for
quaternion numbers). This $\beta$ index in turn identifies the
symmetry group of the ensemble (Orthogonal, Unitary and Symplectic
respectively).

Thanks to the works of Mehta \cite{Mehta} and many others, very
powerful analytical tools are available to deal with invariant
ensembles, both for finite $N$ and as $N\rightarrow\infty$, the
latter limit being usually the most interesting for RM theorists.
However, it was very soon realized that matrices with the smallest
size $N=2$ can equally well provide deep insights and trigger new
ideas, the most successful one being the celebrated Wigner's
surmise \cite{Mehta} which gives an excellent approximation for
the level spacing of bigger matrices. The study of $2\times 2$
random matrices has since been strongly developed and it remains
an active area of research in mathematical physics
\cite{caurier}\cite{lenz}\cite{chau}\cite{ahmed}
\cite{araujo}\cite{kota}\cite{ahmed2}\cite{jackson}\cite{evangelou}\cite{sabbah}\cite{ullah2}\cite{alhassid}
\cite{shaffaf}\cite{vanassche}.

The purpose of the present paper is to introduce and solve exactly
a family of $2\times 2$ random matrices depending on one parameter
$\eta$. This ensemble will have rotational invariance \emph{but} a
real effective Dyson index $\beta_{\mathrm{eff}}$ in an interval.
Although it is commonly assumed that the two properties:
\begin{itemize}
  \item rotational invariance;
  \item real Dyson index.
\end{itemize}
are essentially incompatible, since the Dyson index of an
invariant ensemble is strictly constrained to the values $1,2$ or
$4$ as described above, we will show how to construct explicitly a
counterexample in Section \ref{MainIdea} introducing suitable
correlations among the matrix entries. The motivation for this
study stems from two apparently unrelated issues, namely the
Dumitriu-Edelman $\beta$-ensembles \cite{DE} and the so-called
Poisson-Wigner crossover for the level spacing \cite{BohLes}. In
order to make the paper self-contained, we give a brief
introduction to both of them highlighting also the two main tasks
we tackle in this paper. In subsection \ref{summary}, we provide
the plan of the article.

\subsection{$\beta$-ensembles of Dumitriu-Edelman.}
Consider the jpd \eqref{jpdGauss}. Does there exist a non-trivial
matrix model having \eqref{jpdGauss} as its jpd of eigenvalues for
\emph{any} $\beta>0$? Very recently, Dumitriu and Edelman were
able to answer this question affirmatively \cite{DE}. They
introduced two ensembles of tridiagonal $N\times N$ matrices with
independent entries, whose jpd of eigenvalues is exactly given by
\eqref{jpdGauss} for general $\beta>0$ \cite{DE}. These ensembles
have been called $\beta$-Hermite and $\beta$-Laguerre, according
to the classical weight their jpd contains. This result is
essential for an efficient numerical sampling of random matrices
\cite{vivo} and has triggered a significant amount of further
research
\cite{dumitriu}\cite{forrRain}\cite{killip}\cite{lippert}\cite{esutton}.

Note that the $\beta$-ensembles, having independent non-Gaussian
entries are obviously \emph{non-invariant}. Thus, the first novel
task we tackle in this paper (Section \ref{Gaussian}) is the
following:
\begin{Task}
Design and solve exactly a $(2\times 2)$ ensemble with:
\begin{itemize}
  \item rotational invariance;
  \item running $\beta_{\mathrm{eff}}\geq 0$\myfootnote{Comments on the case $\beta_{\mathrm{eff}}\equiv 0$ are given in
Section \ref{Gaussian}.};
  \item assigned classical potential (in particular,
  Gaussian-Hermite).
\end{itemize}
\end{Task}

In fact, an invariant matrix model displaying a running Dyson
index would be of great interest: tuning the strength of the
correlations between the eigenvalues in \eqref{jpdGauss} has
significant importance for systems which, although endowed with an
intrinsic invariance, are subjected to a weak non-invariant
perturbation (see e.g. \cite{zyc}) and may also have important
implications for lattice gas theory \cite{bakerforr}. Furthermore,
it is a long-standing observation that nuclear systems with
two-body interactions display an average density of states whose
profile is much closer to a Gaussian distribution
\cite{bohigas}\cite{french} than to a semicircle. Hence, a RM
approach with the appropriate symmetries clearly requires much
weaker, and possibly suppressed altogether, correlations among the
energy levels than those arising from \eqref{jpdGauss} with
integer and fixed $\beta$. In this respect, the limit
$\beta_{\mathrm{eff}}\rightarrow 0$ of our model is particularly
appealing (see Section \ref{Gaussian}).

\subsection{Poisson-Wigner crossover.}
Another interesting transitional regime in quantum chaos theory,
namely the so-called Poisson-Wigner crossover for level spacings,
has attracted much attention in the past twenty years
\cite{BohLes}. In terms of the dimensionless nearest-neighbor
spacing $s$, the Poisson and Wigner distributions are given by:
\begin{align}
  P_{\mathrm{POI}}(s) &= \exp(-s) \label{poi}\\
  P_{\mathrm{WIG}}(s) &= \frac{\pi s}{2}\exp\left(-\frac{\pi
  s^2}{4}\right)\label{wig}
\end{align}
and correspond to the limiting cases of classical dynamics, namely
purely regular and completely chaotic. Intermediate regimes
between those two extremes have been intensely investigated (see
\cite{guhr} for a review), and interpolating phenomenological
formulas have been proposed, the most famous being the Brody
\cite{brody} and Berry-Robnik \cite{berry} distributions. The
quest for a deeper understanding of such a crossover has motivated
many proposals of parametrical random matrix models whose level
spacing distribution interpolates between \eqref{poi} and
\eqref{wig}
\cite{cheon}\cite{ullah}\cite{pato}\cite{caurier}\cite{lenz}\cite{chau}.
Normally, the requirement of rotational invariance is the first to
be dropped in those models. The reason is easy to understand: once
this condition is imposed, the Vandermonde-coupling between the
eigenvalues forces the level spacing $P(s)$ to display a term of
the form $\sim s^\beta$ ($\beta=1,2,4$) and thus is very stiff, at
least for small values of the gap $s$. No meaningful crossover
could occur in such models for any standard choice of the
confining potential. This problem would be overcome by an
invariant model with a tunable index $\beta_{\mathrm{eff}}$ and
thus leads to our second unconventional task (Section
\ref{nongaussian}):
\begin{Task}
Design and solve exactly a $(2\times 2)$ ensemble with:
\begin{itemize}
  \item rotational invariance;
  \item running $\beta_{\mathrm{eff}}\geq 0$;
  \item assigned level-spacing profile.
\end{itemize}
\end{Task}

\subsection{Plan of the paper.}\label{summary}

The $2\times 2$ ensemble we are going to introduce in Section
\ref{MainIdea} is completely defined when one assigns:
\begin{itemize}
\item A symmetry group (SG) (Orthogonal, Unitary or Symplectic),
corresponding to real symmetric, hermitian or quaternion self-dual
matrices;
  \item A weight function, to be defined in Section \ref{MainIdea};
  \item A range for the free parameter $\eta$.
\end{itemize}
As far as the SG is concerned, in the present study we will
confine ourselves to hermitian (unitary invariant) matrices,
although generalizations to other SG may be easily derived (see
Section \ref{generalizations}). The Dyson index for this Unitary
ensemble turns out to be
$\beta_{\mathrm{eff}}=\beta-2\eta=2-2\eta$ and for this reason we
call the ensemble \textsf{$\eta$-UE} ($\eta$-Unitary Ensemble).

In Sections \ref{Gaussian} and \ref{nongaussian} we make two
different choices for the combination (weight function $+$ range
for $\eta$) in order to tackle the tasks 1 and 2 described above.
More precisely:
\begin{itemize}
  \item \underline{Section \ref{Gaussian}}: choosing as an example
  a standard Gaussian potential, we design a \textsf{$\eta$-UE}
  ensemble which is essentially a $(2\times 2)$ $\beta$-Hermite model
  \cite{DE} \emph{plus} rotational invariance for $\eta\in [0,1]$. We compute analytically the
  marginal distributions of the correlated entries in
subsection \ref{Marginal distribution of entries.} and we derive
explicitly the spectral properties in \ref{subsectionspectral}.
These results are then checked by numerical diagonalization of
actual \textsf{$\eta$-UE} samples in subsection
\ref{NumericalSimulations}.
  \item \underline{Section \ref{nongaussian}}: choosing as
  limiting cases the Wigner and the Poisson level-spacing profile,
we design a \textsf{$\eta$-UE}
  ensemble whose level spacing interpolates between the two cases
  for the parameter $\eta\in [1/2,1]$. Following the same
  guidelines, it is in principle possible to extend the analysis
  to an arbitrary pre-assigned level-spacing profile
  $\tilde{P}_\eta(s)$.
\end{itemize}

In Section \ref{generalizations} we discuss generalizations of
this model towards different SG, different weight functions and
extended ranges for $\eta$. At that stage, we will make comments
about some emerging features of our model that appear interesting
to be tackled in future
researches.\\
In Section \ref{conclusions}, we first provide a synthetic table
with a comparison of the main features of all the ensembles
considered in this work, and then we add some concluding remarks.
Some technical derivations are also given in the Appendices.

\section{Main idea and the model.}\label{MainIdea}

Let $\mathbf{P}_\eta[\mathcal{X}]\equiv
\mathbf{P}_\eta(x_{11},\ldots,x_{NN})$ be the joint probability
density of the entries for a $N\times N$ random matrix ensemble,
depending on the parameter $\eta$. If the model is required to be
rotationally invariant, as in our case, two facts must be taken
into account:
\begin{enumerate}
  \item Weyl's Lemma holds \cite{Mehta}, so $\mathbf{P}_\eta[\mathcal{X}]$
  can be only a function of the traces of the first $N$ powers of
  $\mathcal{X}$. We highlight this point by writing hereafter
  starred quantities (such as $\mathbf{P}_\eta^\star:=\mathbf{P}_\eta[\mathcal{X}]$) whenever they are meant to be written in terms of
  traces of powers of $\mathcal{X}$.\label{property1}
  \item the jpd of eigenvalues is given by:
\begin{equation}\label{jpdEigenRotationally}
  P_\eta(\lambda_1,\ldots,\lambda_N)\propto \mathbf{P}_\eta^\star\times \prod_{j<k}|\lambda_j-\lambda_k|^\beta
\end{equation}
where the Vandermonde term comes from integrating out the
`angular' variables in the diagonalization $\mathcal{X}\rightarrow
O\Lambda O^{-1}$. In \eqref{jpdEigenRotationally}, the $\beta$
index can take only the values $1,2$ or $4$ according to the SG of
the ensemble (Orthogonal, Unitary or Symplectic respectively).
\label{property2}
\end{enumerate}

We can specialize the properties \ref{property1} and
\ref{property2} to an ensemble of $2\times 2$ unitary invariant
hermitian matrices:
\begin{equation}\label{2x2}
  \mathcal{X}=\begin{pmatrix}
    x & \frac{t+is}{\sqrt{2}} \\
    \frac{t-is}{\sqrt{2}} & y \\
  \end{pmatrix}
\end{equation}
where $x,y,t,s$ are random variables taken from a jpd
$\mathbf{P}_\eta(x,y,t,s)$ and the $1/\sqrt{2}$ factors are
included for later convenience.

In this simplified case, \eqref{jpdEigenRotationally} becomes:
\begin{equation}\label{jpdEigenRotationally2x2}
  P_\eta(\lambda_1,\lambda_2)\propto
  \mathbf{P}_\eta^\star\times |\lambda_2-\lambda_1|^2
\end{equation}
and we choose to write the \textsf{$\eta$-UE} jpd of entries as:
\begin{equation}\label{jpdEntriesTHM}
\mathbf{P}_\eta^\star :=\frac{\mathcal{W}^\star_\eta
}{[\mathcal{V}^\star]^\eta}
\end{equation}
In \eqref{jpdEntriesTHM}, the weight function
$\mathcal{W}^\star_\eta$ is a non-negative, normalizable and
symmetric function of the eigenvalues, expressed in terms of the
traces $(\Tr\mathcal{X},\Tr\mathcal{X}^2)$. It may depend or not
on the parameter $\eta<3/2$.

Now, we define:
\begin{equation}\label{DefOfVstar}
\mathcal{V}^\star = 2\Tr\mathcal{X}^2-(\Tr\mathcal{X})^2
\end{equation}
and it is easy to prove the following identity involving the rhs
of \eqref{DefOfVstar}:
\begin{equation}\label{RepresPowerSums}
    2\Tr\mathcal{X}^2-(\Tr\mathcal{X})^2 = |\lambda_2-\lambda_1|^2
\end{equation}
Through \eqref{RepresPowerSums}, the Vandermonde-squared coupling
has been represented in terms of traces of powers of $\mathcal{X}$
and the jpd of eigenvalues \eqref{jpdEigenRotationally2x2}
becomes:
\begin{equation}\label{jpdEigenRotationally2x2new}
  P_\eta(\lambda_1,\lambda_2)\propto
  \mathcal{W}_\eta^\star\times |\lambda_2-\lambda_1|^{2-2\eta}
\end{equation}
As $\eta\in\mathcal{I}\subset\mathbb{R}$, the effective Dyson
index $\beta_{\mathrm{eff}}=2-2\eta$ assumes real values in an
interval, while the ensemble keeps its rotational invariance
(unlike in the case of $\beta$-ensembles of Dumitriu-Edelman). The
price to pay is that the entries are no longer independent, but
get correlated through the multiple trace-trace interaction term
$\mathcal{V}^\star$.

One may ask whether the crucial identity \eqref{RepresPowerSums}
is just an algebraic accident holding only for $2\times 2$
matrices or it has a deeper origin. In fact,
\eqref{RepresPowerSums} turns out to be a special case of the more
general identity:
\begin{align}\label{Expansion}
 \nonumber \prod_{j<k}(\lambda_j-\lambda_k)^2
  &=\det\mathcal{\mathbf{M}}_N[\mathcal{X}_N]=\\
&=\left|
  \begin{array}{ccccc}
  N & s_1 & s_2 & \cdots & s_{N-1}\\
  s_1 & s_2 & s_3 & \cdots & s_{N}\\
  s_2 & s_3 & s_4 & \cdots & s_{N+1}\\
  \vdots & \vdots &  & \ddots & \\
  s_{N-1} & s_N & s_{N+1} & \cdots & s_{2(N-1)}
  \end{array}
  \right|
\end{align}
where  $s_k=\Tr\mathcal{X}_N^k$ \cite{dunne}. Since the
Vandermonde-squared coupling is a symmetric polynomial in the
eigenvalues, it can be represented on the basis of power sums
\cite{macdonald}, which are nothing but traces of higher order
powers of $\mathcal{X}_N$. The representation \eqref{Expansion}
precisely encodes this change of basis, which is currently used in
the context of the fractional quantum Hall effect but usually
overlooked in RM studies. Note that the general expansion
\eqref{Expansion} can be used in principle to define a $N\times N$
\textsf{$\eta$-UE} model, although any analytical approach appears
very challenging in this case.

From \eqref{jpdEigenRotationally2x2new}, it is clear that
different choices for the weight function and the range
$\mathcal{I}$ for $\eta$ can be combined to achieve a variety of
results. In particular, we are now ready to tackle the first task
described in the Introduction.

\section{First Task: Gaussian weight function.}\label{Gaussian}

Suppose we choose the confining potential to be harmonic
$V(x)=x^2$. It is then sufficient to make the simple choices
$\mathcal{W}_\eta^\star =
\exp\left(-\frac{1}{2}\Tr\mathcal{X}^2\right)$ and $\eta\in [0,1]$
to design a $\beta_{\mathrm{eff}}$-Hermite model
($\beta_{\mathrm{eff}}\in [0,2]$) \cite{DE} \emph{plus} unitary
invariance which we are going to solve exactly. Before doing that,
we make the following important remark:
\begin{Remark}
Unlike in the case of $\beta$-Hermite ensemble, the value
$\beta_{\mathrm{eff}}\equiv 0$ can be actually reached in
\textsf{$\eta$-UE} for $\eta=1$. This corresponds to independent
normally distributed eigenvalues.
\end{Remark}

From \eqref{jpdEntriesTHM} and \eqref{DefOfVstar} we get:
\begin{equation}\label{Petastar}
  \mathbf{P}_\eta^\star =\mathrm{C}_\eta\frac{e^{-\frac{1}{2}\Tr \mathcal{X}^2}}{[2\Tr\mathcal{X}^2-(\Tr\mathcal{X})^2]^\eta}
\end{equation}
where $\mathrm{C}_\eta$ is a normalization constant.

The resulting jpd of eigenvalues \eqref{jpdEigenRotationally2x2}
can be written as:
\begin{equation}\label{Resulting jpd}
  P_\eta(\lambda_1,\lambda_2)=\mathrm{K}_\eta e^{-\frac{1}{2}(\lambda_1^2+\lambda_2^2)}|\lambda_2-\lambda_1|^{2-2\eta}
\end{equation}
where the normalization constant is given by \cite{DE}:
\begin{equation}\label{Keta}
  \mathrm{K}_\eta=\left(\sqrt{\pi}\, 2^{3-2\eta}\, \Gamma\left(\frac{3}{2}-\eta\right)\right)^{-1}
\end{equation}
Note that \eqref{Resulting jpd} is \emph{almost} equivalent to a
$2\times 2$ $\beta_{\mathrm{eff}}$-Hermite jpd (apart from the
actual $\beta_{\mathrm{eff}}=0$ case which is not included there),
although the underlying matrix model is very different in the two
cases. For a related model with complex eigenvalues, see
\cite{callaway}.

The range of variability for $\eta$ is largely arbitrary (see also
subsection \ref{differentrange}). The choice of $[0; 1]$ is
motivated by a nice duality between the limiting cases $\eta=0$
and $\eta=1$ as in the following table:
\begin{table}[htb]
\begin{center}
\begin{tabular}{|c|c|c|}
\hline & $\eta=0$ & $\eta=1$\\
\hline Correlation among Eigenvalues & Strong & Absent\\
\hline Correlation among Entries & Absent & Strong\\
\hline
\end{tabular}
\end{center}
\caption{The jpd of entries factorizes for $\eta=0$ and gives rise
to GUE (Gaussian Unitary Ensemble) with strongly correlated
eigenvalues. On the contrary, for $\eta=1$ the jpd of eigenvalues
factorizes and the eigenvalues become i.i.d. normal variables,
while the entries are strongly correlated.}
\end{table}

Although \eqref{Petastar} defines completely our
\textsf{$\eta$-UE} model, it is instructive to compute
analytically the marginal distributions for the set of correlated
variables $(x,y,t,s)$ for two reasons:
\begin{enumerate}
  \item These results are numerically implemented in subsection
  \ref{NumericalSimulations} to generate and diagonalize actual samples of
  \textsf{$\eta$-UE} matrices. The numerical results will be
  compared with the spectral properties derived in subsection
  \ref{subsectionspectral}.
  \item The marginal distributions deviate smoothly from the GUE factorized
    marginals as $\eta$ departs from zero, and thus they provide quantitative information
    about the onset of correlations between the entries.
\end{enumerate}

\subsection{Marginal
distribution of entries.}\label{Marginal distribution of entries.}
From \eqref{2x2} one has:
\begin{align}
\Tr\mathcal{X} &=x+y\\
\Tr\mathcal{X}^2 &= x^2+y^2+t^2+s^2
\end{align}
so that \eqref{Petastar} implies:
\begin{equation}\label{Petamarginals}
  \mathbf{P}_\eta(x,y,t,s):=\mathrm{C}_\eta\frac{e^{-\frac{1}{2}(x^2+y^2+t^2+s^2)}}{[2(x^2+y^2+t^2+s^2)-(x+y)^2]^\eta}
\end{equation}
The first task is computing the normalization constant
$\mathrm{C}_\eta$, for which the following integral is needed
[\cite{GR} formula 3.382(4)]:
\begin{equation}\label{IntegralFormula}
  \mathcal{I}(\eta,\ell):=\int_0^\infty \frac{r
  e^{-\frac{1}{2}r^2}}{[2r^2+\ell^2]^\eta}dr=2^{-2\eta}e^{\ell^2/4}\Gamma\left(1-\eta,\frac{\ell^2}{4}\right)
\end{equation}
where $\Gamma(x,y)$ is the incomplete Gamma function.

The constant $\mathrm{C}_\eta^{-1}$ is given by:
\begin{equation}\label{C^-1}
\mathrm{C}_\eta^{-1}=\int_{-\infty}^\infty\cdots\int_{-\infty}^\infty
\frac{dx~dy~dt~ds~e^{-\frac{1}{2}(x^2+y^2+t^2+s^2)}}{[2(x^2+y^2+t^2+s^2)-(x+y)^2]^\eta}
\end{equation}
which becomes upon the change to polar coordinates
$(t,s)\rightarrow (r,\theta)$:
\begin{equation}\label{ConstantAfterPolar}
\mathrm{C}_\eta^{-1}=2\pi\int_{-\infty}^\infty\int_{-\infty}^\infty
dx~dy~e^{-\frac{1}{2}(x^2+y^2)}\mathcal{I}(\eta,x-y)
\end{equation}
After further simplifications and the change of variables
$(x+y,x-y)\rightarrow (\phi,\omega)$ the double integral decouples
and we get:
\begin{equation}\label{C_eta}
\mathrm{C}_\eta^{-1}=\pi^{3/2}2^{3-2\eta}\Gamma\left(\frac{3}{2}-\eta\right)
\end{equation}
Now, we can compute the marginal distribution for the variable $x$
following essentially the same steps (hereafter we use the
notation $\mathbf{p}_{k,\eta}$ with $k=1,2,3$ to denote the
marginals with $k$ variables):
\begin{align}\label{MarginalX}
\nonumber\mathbf{p}_{1,\eta}&(x)
=\mathrm{C}_{\eta}\int_{-\infty}^\infty\cdots\int_{-\infty}^\infty
dy~dt~ds~\mathbf{P}_{\eta}(x,y,t,s)=\\
\nonumber &=2\pi\mathrm{C}_{\eta}\frac{d}{dx}\int_0^x
d\mu\int_{-\infty}^\infty
dy~e^{-\frac{1}{2}(\mu^2+y^2)}\mathcal{I}(\eta,y-\mu)=\\
\nonumber &=2^{2-2\eta}\pi
\mathrm{C}_\eta\frac{d}{dx}\int_{-\infty}^\infty
d\omega\Gamma(1-\eta,\omega^2)\int_\omega^{\omega+x}d\phi
e^{-\phi^2}=\\
\nonumber &=2^{2-2\eta}\pi\mathrm{C_\eta}\int_{-\infty}^\infty
d\omega\Gamma(1-\eta,\omega^2)e^{-(x+\omega)^2}=\\
&=\frac{\left(f_\eta\ast
g\right)(x)}{2\sqrt{\pi}\Gamma\left(\frac{3}{2}-\eta\right)}
\end{align}
where in the last passage we have employed the symmetry
$\omega\rightarrow -\omega$ of the integral and the convolution
$(\ast)$ is between the functions:
\begin{equation}\label{f and g}
  f_\eta(y):=\Gamma(1-\eta,y^2)\qquad g(y):=e^{-y^2}
\end{equation}
In the limiting case $\eta\rightarrow 0$, we expect to recover the
pure GUE marginal distribution for the entry $x$, which is simply
a standard Gaussian. Taken into account that $f_{\eta=0}(y)=g(y)$
and $\Gamma(3/2)=\sqrt{\pi}/2$, one gets:
\begin{equation}\label{Limiting eta tende a 0 for p(x)}
\mathbf{p}_{1,\eta=0}(x)=\frac{1}{\pi}\int_{-\infty}^\infty
d\omega
e^{-\omega^2}e^{-(x-\omega)^2}=\frac{e^{-x^2/2}}{\sqrt{2\pi}}
\end{equation}
as it should.\\

\begin{figure}[htb]
\begin{center}
\includegraphics[bb=91 3 322 146,totalheight=0.40\textheight]{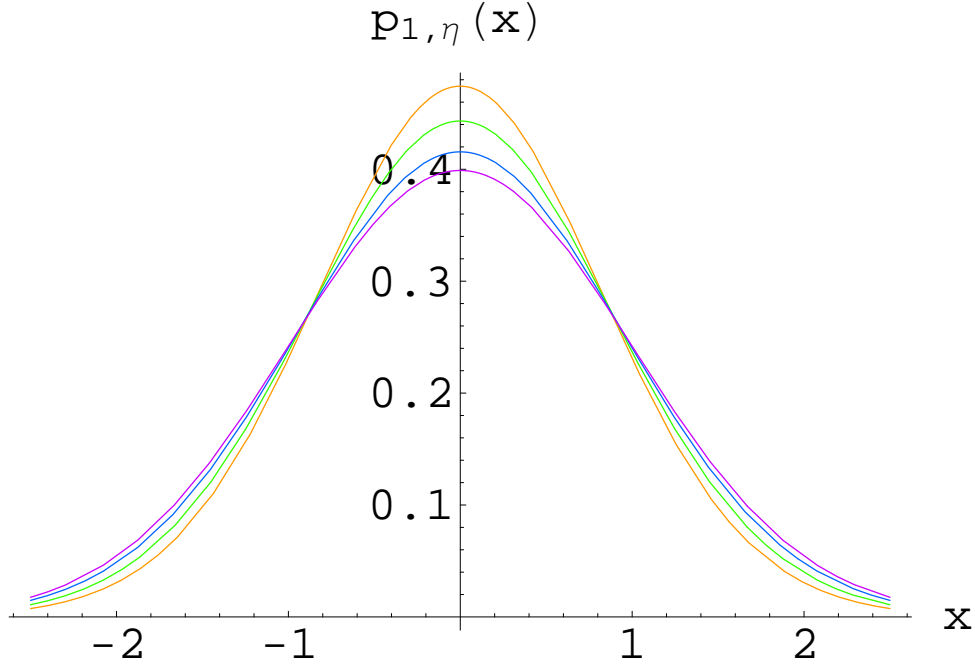} \caption{Plot of the
marginal distribution $\mathbf{p}_{1,\eta}(x)$ for different
values of $\eta$ [Orange $0.8$; Green $0.5$; Blue $0.2$]. The
limiting Gaussian distribution $\eta=0$ is also plotted
[Violet].}\label{marxa}
\end{center}
\end{figure}
A careful asymptotic analysis (see Appendix \ref{Asampt}) of the
convolution integral \eqref{MarginalX} gives for
$x\rightarrow\pm\infty$:
\begin{equation}\label{AsymptoticsMarginalX}
  \mathbf{p}_{1,\eta}(x) \approx \frac{1}{2\,\sqrt{2}\, \Gamma(3/2-\eta)}\,
{\left(\frac{x}{2}\right)}^{-2\eta}\, e^{-x^2/2}
\end{equation}
and for $\eta>0$ the decay is faster than Gaussian due to the
power-law prefactor, in agreement with the plots in figure
\ref{marxa}.

The other marginals can be computed as well:
\begin{align}
\label{othermarg2} \mathbf{p}_{2,\eta} &(x,y)
  =\frac{e^{-\frac{(x+y)^2}{4}}}{4\sqrt{\pi}\Gamma\left(\frac{3}{2}-\eta\right)}\Gamma\left(1-\eta,\frac{(x-y)^2}{4}\right)\\
  \nonumber \mathbf{p}_{3,\eta} &(x,y,t) =
  \frac{e^{-\frac{1}{2}(x^2+y^2+t^2)}}{\pi^{3/2}2^{3-2\eta}\Gamma\left(\frac{3}{2}-\eta\right)}\times\\
  \nonumber &\times\left\{\frac{a^{1/2-\eta}\sqrt{\pi/2}\Gamma(\eta-1/2)~_1
F_1(1/2,3/2-\eta,a/4)}{\Gamma(\eta)}+\right.\\
  &+\left. 2^{1/2-2\eta}\Gamma(1/2-\eta)~_1
  F_1(\eta,1/2+\eta,a/4)\right\}\label{othermarg3}
\end{align}
where we defined $a=(x-y)^2+2 t^2$ and $_1
F_1(\hat{\alpha},\hat{\beta},z)$ is the Kummer confluent
hypergeometric function.

At this stage, we make the following important remarks:

\begin{enumerate}
  \item For $\eta\rightarrow 0$, it is straightforward to check that
\eqref{othermarg2} and \eqref{othermarg3} reproduce the expected
GUE factorized distributions:
\begin{align}\label{ExpectedGUEmarginals}
\mathbf{p}_{2,\eta=0} (x,y) &=\frac{e^{-\frac{1}{2}(x^2+y^2)}}{2\pi}\\
\mathbf{p}_{3,\eta=0} (x,y,t)
&=\frac{e^{-\frac{1}{2}(x^2+y^2+t^2)}}{(2\pi)^{3/2}}
\end{align}
  \item For $\eta\rightarrow 1/2$ (a noticeable special case, see next
  subsection), all the marginal distributions remain well-defined.
  In particular, the apparent divergences of the Gamma functions
  in \eqref{othermarg3} cancel out and the final density reads:
\begin{equation}\label{marg3eta1/2}
\mathbf{p}_{3,\eta=1/2} (x,y,t)=
\frac{1}{\pi^{3/2}2^{5/2}}e^{a/8}K_0(a/8)
\end{equation}
where $K_0(x)$ is a Modified Bessel Function of degree 0 of the
Second Kind.
\end{enumerate}

\subsection{Spectral properties.}\label{subsectionspectral}

As already discussed at the end of the previous subsection, the
case $\eta=1/2$ is particularly interesting, as the jpd of
eigenvalues \eqref{Resulting jpd} collapses onto the Gaussian
Orthogonal Ensemble (GOE) one ($\beta_{\mathrm{eff}}=1$). The
matrices belonging to GOE have \emph{independent} and \emph{real}
entries and the orthogonal group as SG. Instead,
\textsf{$\eta$-UE} reproduces the GOE spectral statistics while
having \emph{complex} and \emph{correlated} entries and the
\emph{unitary} group as \textsc{SG}. This is a first example of a
curious phenomenon, which we call \emph{spectral twinning} (see
Section \ref{differentdyson}) between ensembles having the same
spectral properties (same jpd of eigenvalues) \emph{but} different
\textsc{SG} (different number of independent real variables). As a
consequence, the connection between the exponent of the
Vandermonde $\beta_{\mathrm{eff}}$ and the \textsc{SG} of the
ensemble becomes much more blurred and potentially deceptive than
for the classical $\beta=1,2,4$ ensembles.

In this section and in Appendix \ref{SpecDens}, we compute for
completeness the average spectral density for our
\textsf{$\eta$-UE} model with Gaussian weight function, as this
calculation does not appear to have been carried out explicitly
before. The raw level spacing has already been computed in the
context of a $2\times 2$ $\beta$-Hermite ensemble in
\cite{lecaer}.

The average spectral density $\rho_\eta(\lambda)$ and the gap
probability $P_\eta(s)$ are given as usual by:
\begin{align}
\rho_{\eta}(\lambda) &= \int_{-\infty}^{\infty} P_{\eta}(\lambda,
\lambda_2)\, d\lambda_2\\
P_\eta(s) &= \int_{-\infty}^\infty  P_{\eta}(\lambda_1,
\lambda_1+s)d\lambda_1 \label{foldedgap}
\end{align}
where in \eqref{foldedgap} and hereafter, $s$ is meant as the raw
spacing, without any unfolding procedure being performed on the
spectrum.

They can be computed exactly from \eqref{Resulting jpd} as (see
Appendix \ref{SpecDens}):
\begin{align}
  \label{spectralDensity}\rho_{\eta}(\lambda)&=
\frac{e^{-\lambda^2}}{2^{3/2-\eta}\sqrt{\pi}}\, _1 F_1(3/2-\eta,
1/2, \lambda^2/2)\\
  \label{level spacing}P_\eta(s) &=
  2\sqrt{\pi}\mathrm{K}_\eta~s^{2-2\eta}e^{-s^2/4}
\end{align}
where $\mathrm{K}_\eta$ is defined in \eqref{Keta}.

It is easy to check that \eqref{spectralDensity} recovers for
$\eta=0,1/2,1$ the expected spectral densities (GUE, GOE and
purely Gaussian, respectively):
\begin{align}\label{ExpectedSpectralDensity}
  \rho_0(\lambda) &= \frac{e^{-\lambda^2/2}(1+\lambda^2)}{2\sqrt{2\pi}} \\
  \rho_{1/2}(\lambda) &=
  \frac{e^{-\lambda^2}(2+\sqrt{2\pi}e^{\lambda^2/2}\lambda~\mathrm{erf}(\lambda/\sqrt{2}))}{4\sqrt{\pi}}\\
  \rho_1(\lambda) &= \frac{e^{-\lambda^2/2}}{\sqrt{2\pi}}
\end{align}
where $\mathrm{erf}(z)=(2/\sqrt{\pi})\int_0^z e^{-t^2}dt$.

The results \eqref{spectralDensity} and \eqref{level spacing} are
numerically checked by actual diagonalization of
\textsf{$\eta$-UE} matrices in the next Section.

\subsection{Numerical Simulations.}\label{NumericalSimulations} We
report in this section the results for the spectral density and
the level spacing, obtained by direct sampling of
\textsf{$\eta$-UE} matrices.

The algorithm proceeds as follows: by rejection sampling \cite{NR}
we draw a random number $\bar{x}$ from the marginal distribution
$\mathbf{p}_{1,\eta}(x)$ \eqref{MarginalX}. Then, from the
marginal distribution \eqref{othermarg2}, we determine the
conditional probability:
\begin{equation}\label{ConditionalProbability}
  \mathbf{p}_\eta(y|\bar{x})=\frac{\mathbf{p}_{2,\eta}(\bar{x},y)}{\mathbf{p}_{1,\eta}(\bar{x})}
\end{equation}
and again we draw a random number $\bar{y}$ from
\eqref{ConditionalProbability}. This procedure is iterated through
the higher order marginals up to identifying the four variables
$(\bar{x},\bar{y},\bar{t},\bar{s})$ from which one sample of
\textsf{$\eta$-UE} is constructed. Each sample is then
diagonalized and we give a histogram of the eigenvalues and of the
gaps between the two eigenvalues over a total number of $75000$
samples for each plot.

\begin{figure}[htb]
\begin{center}
\includegraphics[bb=36 189 546 589,totalheight=0.40\textheight]{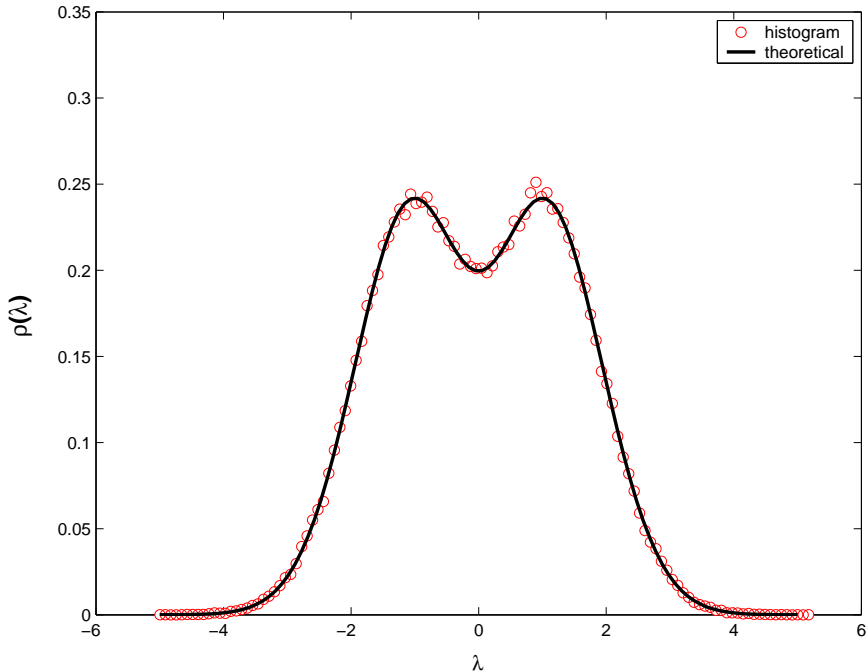}
\caption{Plot of the average density of states
$\rho_\eta(\lambda)$ for $\eta=0$}\label{Specfig1}
\end{center}
\end{figure}

\begin{figure}[htb]
\begin{center}
\includegraphics[bb=36 179 547 589,totalheight=0.40\textheight]{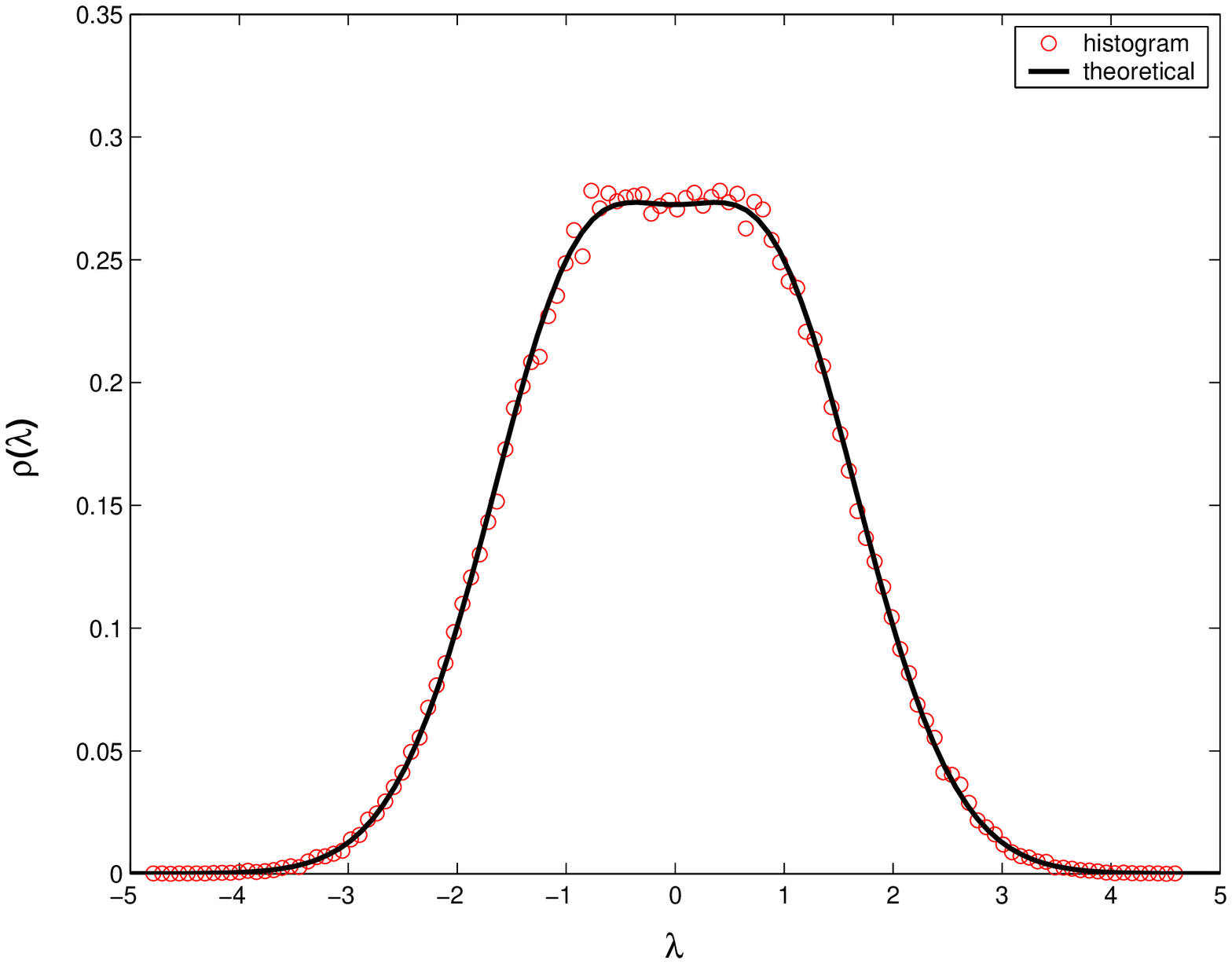}
\caption{Plot of the average density of states
$\rho_\eta(\lambda)$ for $\eta=0.45$}\label{Specfig2}
\end{center}
\end{figure}

\begin{figure}[htb]
\begin{center}
\includegraphics[bb=36 187 546 589,totalheight=0.40\textheight]{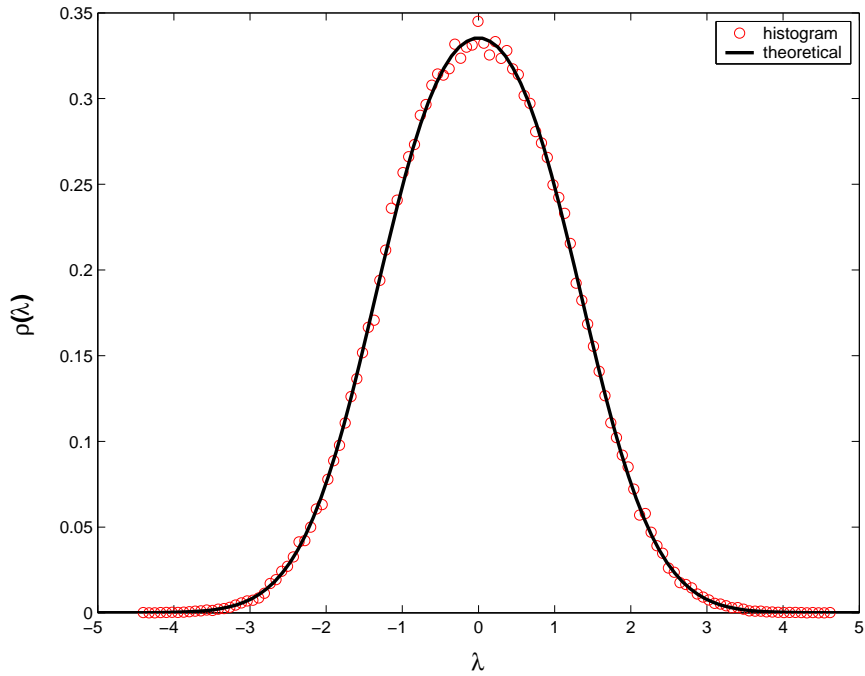}
\caption{Plot of the average density of states
$\rho_\eta(\lambda)$ for $\eta=0.75$. Beyond the value $\eta=0.5$,
the density becomes unimodal.}\label{Specfig3}
\end{center}
\end{figure}

\begin{figure}[htb]
\begin{center}
\includegraphics[bb=49 191 546 589,totalheight=0.40\textheight]{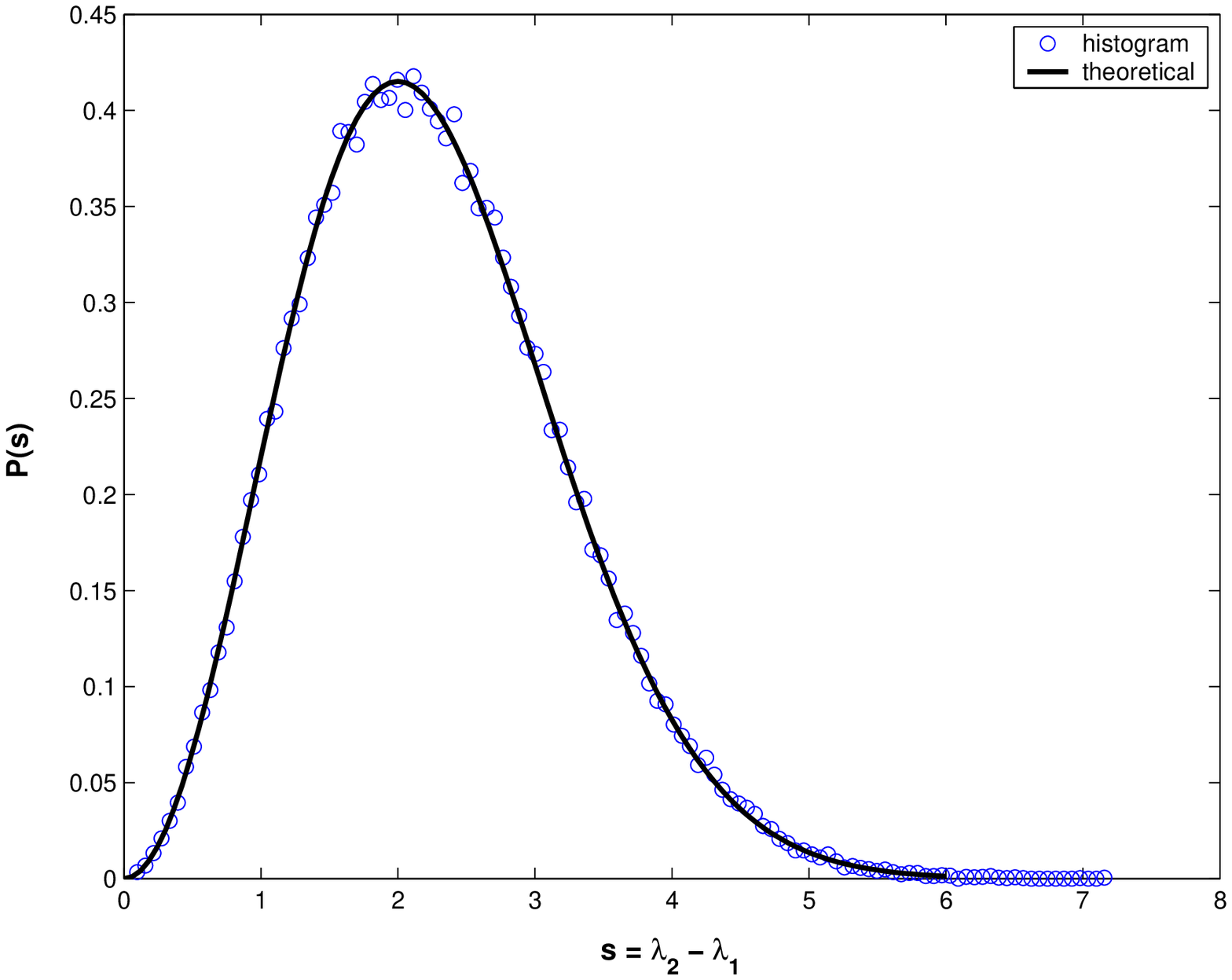}
\caption{Plot of the gap probability $P_\eta(s)$ for
$\eta=0$}\label{Gapfig1}
\end{center}
\end{figure}

\begin{figure}[htb]
\begin{center}
\includegraphics[bb=49 191 546 589,totalheight=0.40\textheight]{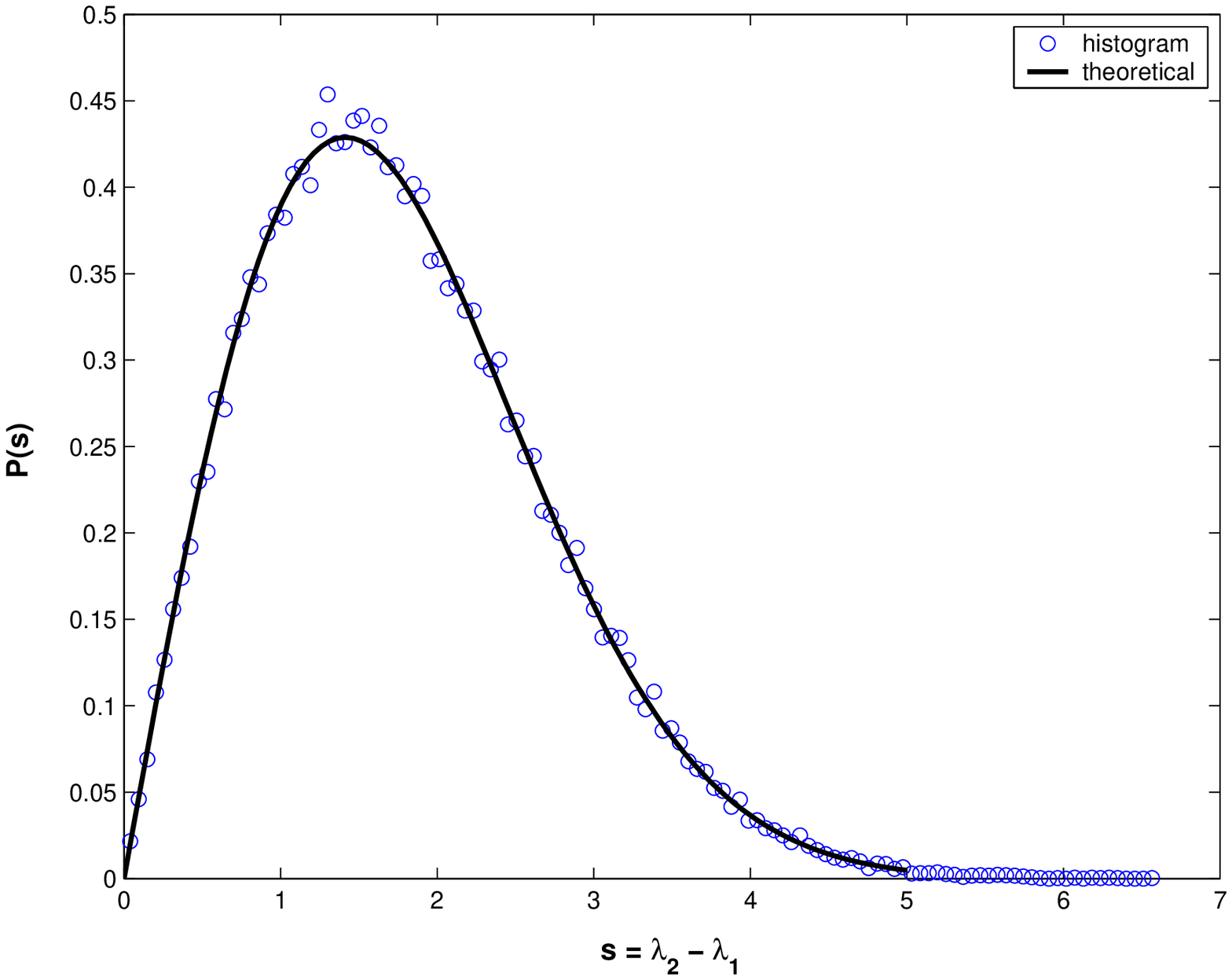}
\caption{Plot of the gap probability $P_\eta(s)$ for
$\eta=0.5$}\label{Gapfig2}
\end{center}
\end{figure}

\begin{figure}[htb]
\begin{center}
\includegraphics[bb=49 184 546 589,totalheight=0.40\textheight]{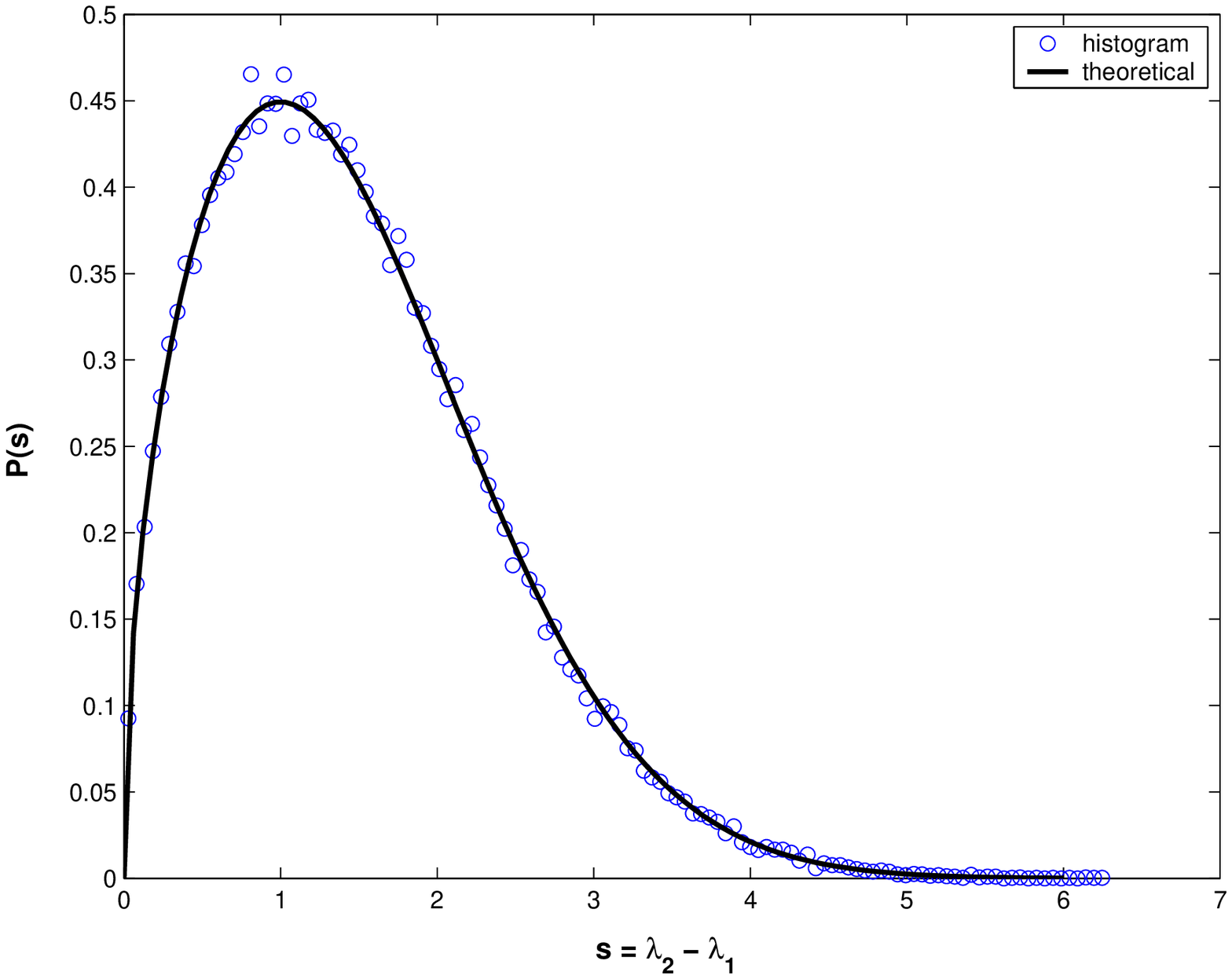}
\caption{Plot of the gap probability $P_\eta(s)$ for
$\eta=0.75$}\label{Gapfig3}
\end{center}
\end{figure}

We include three plots for the spectral density (Fig.
\ref{Specfig1},\ref{Specfig2},\ref{Specfig3}) ($\eta = 0, 0.45,
0.75$ respectively) and three plots for the gap probability (Fig.
\ref{Gapfig1},\ref{Gapfig2},\ref{Gapfig3} ) ($\eta = 0, 0.5, 0.75$
respectively). On top of the histograms, we plot the theoretical
results \eqref{spectralDensity} and \eqref{level spacing}. The
agreement is excellent.

\section{Second Task: Non-Gaussian weight function.}\label{nongaussian}
In this section, we show how to devise a weight function depending
on the parameter $\eta\in [1/2,1]$ such that the level spacing for
the corresponding \textsf{$\eta$-UE} ensemble develops a
Wigner-Poisson crossover. In principle, the solution we offer can
be taken as a guideline for the general problem described in the
Introduction as Task 2. Even though the two above cases (Poisson
and Wigner) absorb the vast majority of literature on these
issues, nevertheless few instances of 'non-standard' gap
distributions have been also reported
\cite{kudo}\cite{pineda}\cite{jakub}. Our hope is that the tool we
propose here may be used to comprise even these fairly anomalous
cases into the universal and otherwise very successful framework
of RM theory.

The starting point is the general jpd of eigenvalues
\eqref{jpdEigenRotationally2x2new}:
\begin{equation}\label{jpdGeneralized}
  P_\eta^{(\mathcal{W})}(\lambda_1,\lambda_2)\propto
  \mathcal{W}^\star_\eta(\lambda_1,\lambda_2)|\lambda_2-\lambda_1|^{2-2\eta}
\end{equation}
where the superscript $(\mathcal{W})$ recalls that the weight
function is still to be determined.

The gap probability for a general weight function can be formally
written as:
\begin{align}\label{Spacing}
  \nonumber P_\eta^{(\mathcal{W})}(&s) =\int_{-\infty}^\infty\int_{-\infty}^\infty~d\lambda_1 d\lambda_2
  \mathcal{W}^\star_\eta(\lambda_1,\lambda_2)|\lambda_2 -
  \lambda_1|^{2-2\eta}\times\\
  &\times\delta(\lambda_2-\lambda_1-s)=s^{2-2\eta}\int_{-\infty}^\infty
  \mathcal{W}^\star_\eta(\lambda_1,\lambda_1+s)d\lambda_1
\end{align}
Apart from being symmetric in the eigenvalues \myfootnote{This is
simply because the joint distribution of eigenvalues can not
depend on how one labels the eigenvalues.}, non-negative
everywhere and normalizable, the sought
$\mathcal{W}^\star_\eta(\lambda_1,\lambda_2)$ should satisfy the
following constraints:
\begin{align}
  \int_{-\infty}^\infty
  \mathcal{W}^\star_{\eta=1}(x,x+s)dx &\propto \exp(-|s|) \label{ReqW1}\\
   \int_{-\infty}^\infty\mathcal{W}^\star_{\eta=1/2}(x,x+s)dx &\propto \exp(-\alpha s^2)\label{ReqW2}
\end{align}
where $\alpha$ is a numerical constant\footnote{The explicit value
for $\alpha$ is $1$ if the spectrum has not been unfolded and
$\pi/4$ in the other case. However, specifying $\alpha$ is not
crucial for what follows.}, in such a way that \eqref{Spacing}
reduces exactly to Poisson for $\eta=1$ or Wigner for $\eta=1/2$.
Indeed, the Wigner distribution in \eqref{wig} corresponds to the
gap probability for a GOE ensemble of $2\times 2$ \emph{real}
matrices ($\beta_{\mathrm{eff}}=1$) and thus has to be realized in
\textsf{$\eta$-UE} for $\eta=1/2$ and for a Gaussian weight
function \eqref{ReqW2}. The other limit is when there are no
correlations among the eigenvalues ($\eta=1$) and thus one may
expect a Poisson distribution for the level spacing \eqref{ReqW1}.

Note that the prefactors in \eqref{ReqW1} and \eqref{ReqW2} can be
restored at the end by normalization and in Eq. (\ref{ReqW1}) we
use the absolute value of $s$ since the function must be symmetric
in $s$.

In order to find an appropriate weight function satisfying the
given constraints, we first make the ansatz\footnote{Other
non-factorized weight functions may exist.}:
\begin{equation}\label{ansatz}
\mathcal{W}^\star_\eta(x,y)=\phi_\eta(x)\phi_\eta(y)
\end{equation}
where $\phi_\eta(x)$ is an even function of $x$.

Now, from \eqref{ReqW1} one has the convolution:
\begin{equation}\label{EvenAnsatz}
 \int_{-\infty}^\infty
  \phi_1(x)\phi_1(s-x)dx = \exp(-|s|)
\end{equation}
which gives in Fourier space:
\begin{equation}\label{Fourier}
  [\tilde{\phi}_1(k)]^2=\frac{2}{k^2+1}
\end{equation}
Thus:
\begin{equation}\label{FourierSqrt}
  \tilde{\phi}_1(k)=\sqrt{\frac{2}{k^2+1}}
\end{equation}
Inverting \eqref{FourierSqrt}, we get:
\begin{equation}\label{InvertFourierSqrt}
\phi_1(x)=\frac{\sqrt{2}}{\pi}K_0(x)
\end{equation}
where $K_0(x)$ is a Modified Bessel Function of degree $0$ of the
Second Kind.

Similarly, for $\eta=1/2$ one gets:
\begin{equation}\label{Eta=1/2}
  \phi_{1/2}(x) = [4\alpha/\pi]^{1/4} e^{-2\alpha x^2}
\end{equation}

In order to obtain the function interpolating between
\eqref{InvertFourierSqrt} and \eqref{Eta=1/2}, we notice that
$K_0(x)$ has the following integral representation \cite{GR}:
\begin{equation}\label{IntRepForK0}
  K_0(x)=\frac{1}{2}\int_0^\infty dt\frac{e^{-\gamma t- x^2/(4\gamma t)} }{t}.
\end{equation}
where $\gamma$ is any real and positive parameter.

Then one can write:
\begin{equation}
\phi_1(x) = \mathrm{B}_1\,\int_0^\infty
\frac{dt}{t}\exp\left(-\gamma t- \frac{x^2}{4\gamma
t}\right)\label{intphi1}
\end{equation}
where $\mathrm{B}_1=1/(\pi \sqrt{2})$. On the other hand,
$\phi_{1/2}(x)$ can also be written trivially in a similar
integral representation as:
\begin{equation}
\phi_{1/2}(x) = \mathrm{B}_{1/2} \int_0^{\infty}
\exp\left(-\bar{\gamma} t - x^2/{4\bar{\gamma}}\right) dt
\label{intphi1/2}
\end{equation}
where $\mathrm{B}_{1/2}= 2^{-5/2} \alpha^{-3/4} {\pi}^{-1/4}$ and
$\bar{\gamma}:=1/(8\alpha)$ . Thus, for general $1/2\le \eta\le
1$, a natural definition of $\phi_{\eta}(x)$ interpolating the two
bordering cases would be:
\begin{equation}
\phi_{\eta}(x) = \mathrm{B}_{\eta} \int_0^{\infty}
t^{1-2\eta}\exp\left(-\frac{t}{8\alpha}-\frac{2\alpha
x^2}{t^{2\eta-1}}\right) \, dt. \label{intphi}
\end{equation}
The corresponding weight function $\mathcal{W}^\star_\eta(x,y)$ is
given by the product in \eqref{ansatz} and satisfies all the given
constraints. We will refer to this weight function as a
Generalized Bessel weight.

The reader may be puzzled by the non-standard expression
\eqref{intphi} and may wonder whether the resulting weight
function $\mathcal{W}^\star_\eta(\lambda_1,\lambda_2)$ is indeed
expressible in terms of traces of powers of $\mathcal{X}$, a fact
which is not obvious at first sight. Actually, this can be shown
easily by expressing the individual eigenvalues as:
\begin{equation}\label{Eigenvaluestraces}
  \lambda_{1,2}=\frac{s_1\pm\sqrt{2 s_2 -s_1^2}}{2}
\end{equation}
where $s_j = \Tr\mathcal{X}^j$.

The gap probability can then be computed from \eqref{Spacing}. For
arbitrary $\eta$, the integral in \eqref{Spacing} is difficult to
carry out for all $s$. However, the large $s$ tail of
$P_{\eta}^{(\mathcal{W})}(s)$ can be easily derived by the saddle
point method and has the following behavior:
\begin{equation}
P_{\eta}^{(\mathcal{W})}(s) \propto \exp\left[-\eta
(2\eta-1)^{-1+1/{2\eta}}\, (2\alpha)^{-1+1/\eta}\,
s^{1/\eta}\right]. \label{gapasymp}
\end{equation}
One can check easily that it reduces to the known Wigner and
Poisson cases for $\eta=1/2$ and $\eta=1$ respectively. Thus our
ensemble smoothly interpolates the tail of the gap distribution
between the two known cases, although the full behavior of
$P_{\eta}^{(\mathcal{W})}(s)$ is different from any previous
proposals (e.g. Brody and Berry-Robnik distributions).

\begin{figure}[htb]
\begin{center}
\includegraphics[bb=91.5625 3.1875 321.938 145.5,totalheight=0.40\textheight]{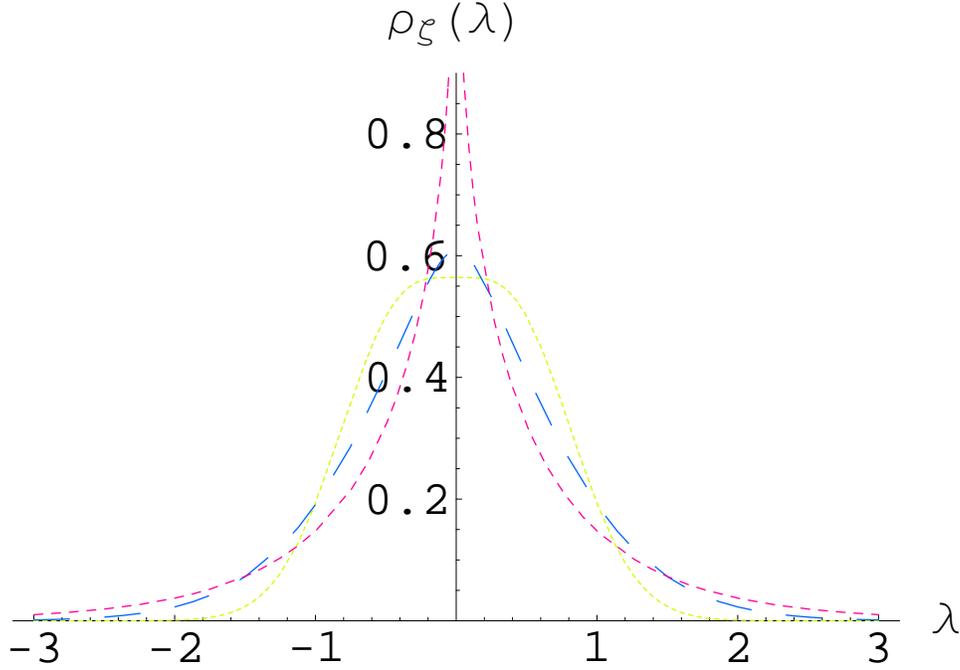} \caption{Plot of the
spectral density $\rho_\zeta(\lambda)$ for the following different
values of $\zeta$: 0 (dotted green), 0.45 (long-dashed blue), 0.9
(short-dashed magenta). Note the pure GOE density for $\zeta=0$
and the peculiar trend towards an integrable divergence at the
origin as $\zeta\rightarrow 1^{-}$.}\label{RhoNonGaussian}
\end{center}
\end{figure}

Also the spectral density can be numerically investigated after
some intermediate algebraic steps that we include in Appendix
\ref{nongaussappendix} for completeness. The resulting plots for
different values of $\zeta:=2\eta-1$ ($0\leq\zeta\leq 1$) and
$\alpha=1$ are reported in Fig. \ref{RhoNonGaussian}.

In summary, the Generalized Bessel weight function
$\mathcal{W}^\star_\eta(\lambda_1,\lambda_2)=\phi_\eta(\lambda_1)\phi_\eta(\lambda_2)$,
where $\phi_\eta(x)$ is given by \eqref{intphi} and
$\lambda_{1,2}$ are expressed in terms of traces as
\eqref{Eigenvaluestraces} generates a \textsf{$\eta$-UE} model
having:
\begin{itemize}
  \item rotational invariance;
  \item $0\leq\beta_{\mathrm{eff}}\leq 1$;
  \item prescribed level spacing (interpolating between Poisson and
  Wigner);
  \item a novel transitional profile for the spectral density as documented in Fig. \ref{RhoNonGaussian}.
\end{itemize}
and is to be regarded as complementary to the model proposed in
\cite{chau}. Note also that the above results are still valid in
the range $[0,1/2]$ for $\eta$, corresponding to a GUE-Poisson
crossover.

\section{Generalizations.}\label{generalizations}
The work presented here can be extended in several directions. We
would like to offer a list of issues that can be tackled in future
researches.

\subsection{Different Dyson class $\beta$.}\label{differentdyson}
We confined our investigation to hermitian matrices $(\beta=2)$,
since the identity \eqref{RepresPowerSums} involves exactly the
exponent $2$ for the Vandermonde, but it is not harmful to
consider the following obvious relations instead:
\begin{align}\label{obvious relation}
|\lambda_2-\lambda_1| &= \sqrt{2\Tr\mathcal{X}^2-(\Tr\mathcal{X})^2} \\
|\lambda_2-\lambda_1|^4 &=
\left[2\Tr\mathcal{X}^2-(\Tr\mathcal{X})^2\right]^2
\end{align}
and reformulate the model for real symmetric and quaternion
self-dual Hermitian matrices respectively:
\begin{equation}\label{2x2real}
  \mathcal{X}_{\mathbb{R}}=\begin{pmatrix}
    x & t \\
    t & y \\
  \end{pmatrix}\qquad
\mathcal{X}_{\mathbb{H}}=\left(\begin{array}{cc|cc}
x & 0 & t+is & u+iv \\
0 & x & -u+iv & t-is\\
\hline t-is & -u-iv & y & 0\\
u-iv & t+is & 0 & y
\end{array}\right)
\end{equation}
having respectively $3$ and $6$ real independent
variables.\\
\\
\textsc{Example:} real symmetric matrices with Gaussian weight
function. In this case, the jpd of eigenvalues reads:
\begin{equation}\label{jpd eigenvalues real Gauss}
P_{\hat{\eta}}(\lambda_1,\lambda_2)\propto
  \exp\left(-\frac{1}{2}(\lambda_1^2+\lambda_2^2)\right)
  |\lambda_2-\lambda_1|^{1-\hat{\eta}}
\end{equation}
where we rename the running parameter as $0\leq\hat{\eta}\leq 1$,
and the jpd of entries is given by:
\begin{equation}\label{PetastarrealGaussian}
  \mathbf{P}_{\hat{\eta}}^\star =\mathrm{C}_{\hat{\eta}}\frac{e^{-\frac{1}{2}\Tr \mathcal{X}_{\mathbb{R}}^2}}
  {[2\Tr\mathcal{X}_{\mathbb{R}}^2-(\Tr\mathcal{X}_{\mathbb{R}})^2]^{\hat{\eta}/2}}
\end{equation}
The expression \eqref{jpd eigenvalues real Gauss}, when compared
with the corresponding one for hermitian matrices \eqref{Resulting
jpd}, leads immediately to the following observation: if the
parameter $\eta$ is chosen in $[1/2,1]$ and $\hat{\eta} :=
2\eta-1$, then the two ensembles (real and complex) get twinned,
i.e. they share the same spectral properties, despite belonging to
different classes of invariance and having even a different number
of independent variables. This \emph{spectral twinning} is a
curious byproduct of our construction, which was already remarked
in subsection \ref{subsectionspectral}. While it is premature to
imagine possible physical application for this, nonetheless we
believe that this peculiar property, which does not hold for any
classical invariant ensemble and of course neither for the
$\beta$-ensembles, deserves further investigations and may be
related to some group-theoretical features of \textsf{$\eta$-XE}
(\textsf{X}=\textsf{O},\textsf{U},\textsf{S}) yet undiscovered.
Note also that this property would survive even for $N\times N$
\textsf{$\eta$-XE} and holds for any acceptable weight function.

\subsection{Different classical weights.}\label{differentweights}
A whole group of novel $2\times 2$ ensembles can be generated by
choosing different weight functions among the classical ones, as:
\begin{enumerate}
  \item  Laguerre and Jacobi (to make contact with
\cite{DE});
  \item fixed and restricted trace ensembles \cite{rosen}\cite{akemann}\cite{lecaer2};
  \item quartic and higher order potentials
  \cite{Brezin};\label{quartic}
  \item power laws
  \cite{bertuola}\cite{toscano}\cite{abulmagd};\label{powerlaw}
\end{enumerate}
In view of subsection \ref{differentdyson}, all the above can be
generated starting from real, complex and quaternion entries and a
number of twinnings can be found. In particular, the
$\beta$-version of (\ref{quartic}) and (\ref{powerlaw}) has not
been constructed as an actual random matrix ensemble so far, while
this problem can be tackled using the method we presented here.

\subsection{Extended range for $\eta$.}\label{differentrange}
Provided that the weight function decays fast enough at infinity
in order to ensure convergence of the integrals involved, it is
possible to extend considerably the range of variability for
$\eta$ in the two examples above as well as in any future study.
Taken the Gaussian case as an example (Section \ref{Gaussian}),
the following extensions can be considered:
\begin{itemize}
  \item $\eta <0$: a negative $\eta$ in \eqref{Resulting jpd} enhances
(instead of suppressing) the correlations among eigenvalues and
extends the range of variability for the effective Dyson index
from $[0,2]$ to $[0,+\infty)$.
  \item $1<\eta<3/2$: this case is even more interesting as it
  allows to generate an \textsf{Anti-$\eta$-UE} ensemble whose jpd of
  eigenvalues would be given by:
\begin{equation}\label{AntiGUE}
  P(\lambda_1,\lambda_2):=\tilde{\mathrm{K}}_\eta \exp\left(-\frac{1}{2}(\lambda_1^2+\lambda_2^2)\right)
|\lambda_2-\lambda_1|^{-\beta_{\mathrm{eff}}}\qquad
0<\beta_{\mathrm{eff}}<1
\end{equation}
i.e. with a \emph{negative} Dyson index. The most immediate
consequence is that the peculiar level repulsion has to be
replaced by a fairly uncommon (at least in RM studies) \emph{level
attraction}. The tendency of energy levels to cluster instead of
repelling apart has been found in several disordered many-body
systems \cite{hsu2}\cite{jalab}\cite{skv}\cite{bolte} but has not
received equal attention in the context of invariant RM. The
\textsf{Anti-$\eta$-UE} is likely to lead to a
'non-Wigner'-surmise for the level spacing to be compared with the
studies cited above. We offer this idea as our last contribution
in this paper.
\end{itemize}

\section{Conclusions.}\label{conclusions}
Before summarizing the main results of this paper, we propose the
synthetic Table II, containing the most relevant features of the
ensembles considered in this work.\\
\begin{table}[htb]
\begin{center}
\begin{tabular}{|c|c|c|c|c|c|c|}
\hline \textbf{Name} & \textbf{Weight} & \textbf{Inv.} &
\textbf{Indep.} & \textbf{Entries} & \textbf{Size}
& \textbf{$\beta_{\mathrm{eff}}$}\\
\hline \hline
 $\beta$-Hermite & Gaussian & N & Y & $\mathbb{R}$ & $N\geq 2$ & $>0$\\
 \hline
 GUE & Gaussian & Y & Y & $\mathbb{C}$ & $N\geq 2$ & $2$ \\
 \hline
  GOE & Gaussian & Y & Y & $\mathbb{R}$ & $N\geq 2$ & $1$\\
  \hline
\textsf{$\eta$-UE} & Gaussian & Y & N & $\mathbb{C}$ & $N=2$ & $[0,2]$\\
\cline{2-7}
 & Generalized Bessel & Y & N & $\mathbb{C}$ & $N=2$ & $[0,1]$\\
 \hline
 \textsf{Anti-$\eta$-UE} & Gaussian & Y & N
 & $\mathbb{C}$ & $N=2$ & $[-1,0]$\\
\hline\hline \textsf{$\beta$-XE}\myfootnotemark &
Arbitrary\myfootnotemark
& Y & N & $\mathbb{R},\mathbb{C},\mathbb{H}$ & $N=2$ & Arbitrary\myfootnotemark\\
\hline
\end{tabular}
\end{center}\vspace{5pt}
\caption{Columnwise: Name of the ensemble, Classical weight or
weight function, Rotational Invariance, Independent Entries, Type
of Entries, Size of the matrices, Range for $\beta_{\mathrm{eff}}$
(exponent of the Vandermonde). In the last row, different
combinations of symmetry class, weight function and range for
$\eta$ may be exploited according to Section
\ref{generalizations}.}
\end{table}
\myfootnotetext{See Section \ref{differentdyson}}
\myfootnotetext{See Section \ref{differentweights}}
\myfootnotetext{See Section \ref{differentrange}}

The main results of the paper can be summarized as follows:
\begin{enumerate}
  \item Although the $\beta$-index of an invariant ensemble is
  determined uniquely by its symmetry group, through a rather
  unusual expansion of the Vandermonde-squared
on the basis of power sums it is possible to neutralize (entirely
or partially) the coupling between the eigenvalues introducing
suitable correlations among the entries. Hence, one has to be
careful in deducing the SG of the ensemble from the exponent of
the Vandermonde coupling $\beta_{\mathrm{eff}}$, as this
connection may be deceptive.
  \item Using this tool, we have constructed an invariant $2\times 2$ version of the $\beta$-ensembles of Dumitriu
  and Edelman. This matrix model is
  completely defined assigning a Symmetry Group (Orthogonal, Unitary,
  Symplectic), a weight function and a certain range
  for $\eta$:
\begin{itemize}
  \item For the Unitary case, with Gaussian weight function and $0\leq\eta\leq 1$,
  both the marginal distribution of the entries and the spectral properties have
  been computed analytically and have
  been tested by numerical sampling of \textsf{$\eta$-UE}
  matrices. The case $\beta_{\mathrm{eff}}=0$, corresponding to
  independent normally distributed eigenvalues, is particularly
  interesting and can be obtained for $\eta=1$.
  \item For the Unitary case, with a Generalized Bessel weight function and $1/2\leq\eta\leq 1$,
  we generate \textsf{$\eta$-UE} matrices with a level spacing profile
  interpolating between Poisson and Wigner. Both analytical and
  numerical results have been provided.
  \end{itemize}
  \item Unlike the classical invariant ensembles,
  our ensembles belonging to different symmetry classes
  may display the same spectral properties for a suitable range of
  the free parameter. We called this curious phenomenon \emph{spectral
  twinning} and we leave a deeper understanding of it as an open
  problem.
  \item An extended range for $\eta$ may lead to invariant
  ensembles whose eigenvalues tend to cluster instead of repelling
  apart due to a negative Dyson index $\beta$. We are not aware of
  any previous proposal in this direction, even though this
  behavior is fairly common in the study of disordered many-body
  systems.
\end{enumerate}
While generalizations to bigger sizes $N>2$ appear difficult to be
tackled analytically, nonetheless the model we presented here
displays non-trivial and often surprising features, which make us
hope that the proposed technical tool may prove useful in future
RM studies.

\section*{Acknowledgments.}
PV has been supported by a Marie Curie Early Stage Training
Fellowship (NET-ACE project). We are grateful to Oriol Bohigas for
his constant advice and support and to Gernot Akemann, Giovanni
Cicuta and Leonid Shifrin for helpful discussions and comments. We
also thank Elisa Garimberti for a careful revision of the
manuscript.

\appendix
\section{Asymptotic Analysis of $ \mathbf{p}_{1,\eta}(x)$.}\label{Asampt}
We start from the exact expression of the marginal distribution
\eqref{MarginalX}:
\begin{equation}
 \mathbf{p}_{1,\eta}(x)= \frac{1}{2\sqrt{\pi} \Gamma(3/2-\eta)}\,
\int_{-\infty}^{\infty} dy \Gamma(1-\eta, y^2)\, e^{-(x+y)^2}
\label{mar1}
\end{equation}
We first divide the integral into 2 parts: over $[-\infty,0]$ and
$[0,\infty]$. In the first part, we also make a change of variable
$y\to -y$. This gives:
\begin{align}
\nonumber  \mathbf{p}_{1,\eta}(x) &= \frac{1}{2\sqrt{\pi}
\Gamma(3/2-\eta)}\,\left[\int_0^{\infty} dy \Gamma(1-\eta, y^2)\,
e^{-(x-y)^2} + \right.\\
\nonumber &\left. +\int_0^{\infty} dy \Gamma(1-\eta, y^2)
e^{-(x+y)^2}\right]= \\
&= \frac{1}{2\sqrt{\pi}
\Gamma(3/2-\eta)}\left[J_1(x)+J_2(x)\right] \label{mar2}
\end{align}

Consider first $J_1(x)$. Clearly, the most important contribution
to the integral comes from the region around $y=x$ for large $x$.
Since $y$ is large in this regime, we can replace $\Gamma(1-\eta,
y^2)$ by its leading asymptotic form:
\begin{equation}
\Gamma(1-\eta, y^2)\approx e^{-y^2}\, y^{-2\eta}. \label{asymp1}
\end{equation}
Substituting this in the expression for $J_1(x)$ we get, to
leading order:
\begin{equation}
J_1(x) \approx \int_{\bar{a}}^{\infty} dy\, e^{-y^2}\,
y^{-2\eta}\, e^{-(x-y)^2} \label{asymp2}
\end{equation}
where the lower limit $\bar{a}\sim O(1)$. Rewriting the
exponential we get:
\begin{equation}
J_1(x) \approx e^{-x^2/2} \int_{\bar{a}}^{\infty}\, dy\,
y^{-2\eta}\, e^{-2(y-x/2)^2} \label{asymp3}
\end{equation}
Making a change of variable, $\sqrt{2}\,(y-x/2)=u$, we get for
large $x$:
\begin{equation}
J_1(x)\approx \frac{1}{\sqrt{2}}\, e^{-x^2/2}\, (x/2)^{-2\eta}\,
\int_{\sqrt{2}(\bar{a}-x/2)}^{\infty} du\, e^{-u^2} \label{asymp4}
\end{equation}
In the $x\to \infty$ limit, the lower limit of the integral can be
replaced by $-\infty$. This finally gives:
\begin{equation}
J_1(x) \approx \sqrt{\frac{\pi}{2}} \, e^{-x^2/2}\, (x/2)^{-2\eta}
\label{asymp5}
\end{equation}

Now for the integral $J_2(x)$, for large $x$, the most important
contribution comes from the region $y=0$. Thus, it is easy to see
that for large $x$, $J_2(x) \approx e^{-x^2} \int_0^{\infty} dy
\Gamma(1-\eta, y^2)= b_\eta e^{-x^2}$ where $b_\eta$ is a
constant. Thus, clearly $J_2(x)$ decays faster than $J_1(x)$ and
hence can be neglected for large $x$.

Thus, for large $x$, to leading order we get:
\begin{equation}
 \mathbf{p}_{1,\eta}(x)\approx \frac{1}{2\,\sqrt{2}\, \Gamma(3/2-\eta)}\,
{\left(\frac{x}{2}\right)}^{-2\eta}\, e^{-x^2/2} \label{asympf}
\end{equation}
It is easy to check that for $\eta=0$, \eqref{asympf} reduces to
the standard Gaussian.
\section{Spectral Density for the Gaussian weight function.}\label{SpecDens}

Consider the joint distribution of the two eigenvalues
\eqref{Resulting jpd}:
\begin{equation}
P_{\eta}(\lambda_1,\lambda_2)= \mathrm{K}_\eta\,
e^{-(\lambda_1^2+\lambda_2^2)/2}\, |\lambda_2-\lambda_1|^{2-2\eta}
\label{jpdf}
\end{equation}
where the normalization constant $\mathrm{K}_\eta$ is given by
\eqref{Keta}.

We want to compute the average density of states which is
precisely the marginal distribution:
\begin{equation}
\rho_{\eta}(\lambda)= \int_{-\infty}^{\infty} P_{\eta}(\lambda,
\lambda_2)\, d\lambda_2 \label{marg1}
\end{equation}
Substituting the jpd from \eqref{jpdf} into \eqref{marg1} and
making a change of variable $\lambda_2-\lambda=x$ we get:
\begin{equation}
\rho_{\eta}(\lambda)= \mathrm{K}_\eta\, e^{-\lambda^2/2}\,
\int_{-\infty}^{\infty} e^{-(\lambda+x)^2/2}\, |x|^{2-2\eta}\, dx
\label{q1}
\end{equation}
Divide the integral into two parts: over $[-\infty,0]$ and
$[0,\infty]$ and write it as:
\begin{align}
\nonumber\rho_{\eta}(\lambda) &= \mathrm{K}_\eta\,
e^{-\lambda^2/2}\, \left[\int_0^{\infty}
\left\{e^{-(\lambda-x)^2/2} +
e^{-(\lambda+x)^2/2}\right\}\times\right.\\
&\left.\times \,x^{2-2\eta}\, dx\right]= \mathrm{K}_\eta\,
e^{-\lambda^2/2}\,\left[I_1(\lambda)+I_2(\lambda)\right]
\label{q2}
\end{align}
The first integral can be rewritten as:
\begin{align}
\nonumber I_1(\lambda)&= e^{-\lambda^2}\,\int_0^{\infty}\,
e^{x\lambda -x^2/2}\, x^{2-2\eta}\, dx = e^{-\lambda^2/4}\,
\Gamma(3-2\eta)\,\times\\
&\times D_{2\eta-3}(-\lambda) \label{d1}
\end{align}
where $D_p(z)$ is the parabolic cylinder function of index $p$ and
argument $z$ \cite{GR}. Similarly $I_2(\lambda)= I_1(-\lambda)$.
Adding the two we get:
\begin{equation}
\rho_{\eta}(\lambda)= \mathrm{K}_\eta\,\Gamma(3-2\eta)\,
e^{-3\lambda^2/4}\,\left[D_{2\eta-3}(\lambda)+
D_{2\eta-3}(-\lambda)\right] \label{marg2}
\end{equation}

We next use the identity (9.240) of \cite{GR}:
\begin{equation}
D_p(z) +D_p(-z)=
\frac{2^{p/2+1}\sqrt{\pi}e^{-z^2/4}}{\Gamma((1-p)/2)}\, \,
_1F_1(-p/2, 1/2, z^2/2) \label{iden1}
\end{equation}
where $ _1 F_1(\hat{\alpha},\hat{\beta},z)$ is the Kummer
confluent hypergeometric function defined by the series:
\begin{equation}
_1 F_1(\hat{\alpha},\hat{\beta},z)= 1+
\frac{\hat{\alpha}}{\hat{\beta}} \frac{z}{1!}+
\frac{\hat{\alpha}(\hat{\alpha}+1)}{\hat{\beta}(\hat{\beta}+1)}\frac{z^2}{2!}+\ldots
\label{dhyper}
\end{equation}
Using this identity in \eqref{marg2} we get:
\begin{align}
\nonumber\rho_{\eta}(\lambda) &= \frac{
\mathrm{K}_\eta\,2^{\eta-1/2}\,
\sqrt{\pi}\,\Gamma(3-2\eta)}{\Gamma(2-\eta)}\,
e^{-\lambda^2}\,\times\\
&\times~_1F_1 (3/2-\eta, 1/2, \lambda^2/2) \label{marg3}
\end{align}

We then use the explicit expression of $\mathrm{K}_{\eta}$ from
\eqref{Keta} to get:
\begin{equation}
\rho_{\eta}(\lambda)= \,
\frac{2^{3\eta-7/2}\Gamma(3-2\eta)}{\Gamma(3/2-\eta)\Gamma(2-\eta)}\,
e^{-\lambda^2}\, _1 F_1(3/2-\eta, 1/2, \lambda^2/2) \label{marg4}
\end{equation}
We can further simplify the constant term by using the Gamma
function identity (doubling formula):
\begin{equation}
\Gamma(2z)= \frac{2^{2z-1}}{\sqrt{\pi}}\, \Gamma(z)\,
\Gamma(z+1/2) \label{gamma1}
\end{equation}
This then gives our final formula:
\begin{equation}
\rho_{\eta}(\lambda)=
\frac{e^{-\lambda^2}}{2^{3/2-\eta}\sqrt{\pi}}\, _1 F_1(3/2-\eta,
1/2, \lambda^2/2) \label{marg5}
\end{equation}

\section{Spectral Density for the Generalized Bessel weight function.}\label{nongaussappendix}

The spectral density for the Generalized Bessel weight function
\eqref{ansatz} is given by:
\begin{equation}\label{specdensnongauss}
  \rho_\zeta(\lambda)\propto \phi_\zeta(\lambda)
  \int_{-\infty}^\infty dy~ \phi_\zeta(y)|\lambda-y|^{1-\zeta}
\end{equation}
where the proportionality constant is just given by the
normalization condition and we have slightly changed the notation
$\phi_\eta(x)\rightarrow\phi_\zeta(x)$, where:
\begin{equation}\label{phizeta}
  \phi_\zeta(x):=\int_0^\infty dt~t^{-\zeta}\exp\left(-\frac{t}{8\alpha}-2\alpha x^2 t^{-\zeta}\right)
\end{equation}
and $0\leq\zeta\leq 1$.

We aim to rewrite \eqref{specdensnongauss} in a form suitable for
a numerical implementation, so from now on we will drop all the
constant in front and we will normalize the numerical density to
$1$ at the very end.

First, we define:
\begin{equation}\label{F}
  F_\zeta(x):=\int_0^\infty ds~s^{1-\zeta}\phi_\zeta(s+x)
\end{equation}
and it is easy to check that:
\begin{equation}\label{rhoaftermanysubs}
\rho_\zeta(\lambda)\propto \phi_\zeta(\lambda)
  \left[F_\zeta(\lambda)+F_\zeta(-\lambda)\right]
\end{equation}
Exploiting \eqref{phizeta} and changing the order of integration,
$F_\zeta(x)$ can be rewritten as:
\begin{equation}\label{Fxaftersubs}
F_\zeta(x)\propto \int_0^\infty dt~t^{-\zeta}
\exp\left(-\frac{t}{8\alpha}-2\alpha x^2 t^{-\zeta}\right)
\int_0^\infty ds~s^{1-\zeta}\exp\left(-\varpi s^2-\kappa s\right)
\end{equation}
where $\varpi:=2\alpha t^{-\zeta}$ and $\kappa:=4\alpha x t
^{-\zeta} $.

The $s$-integral can be carried out analytically and thanks again
to the identity (9.240) of \cite{GR} we get:
\begin{align}\label{FinalResultForF}
\nonumber F_\zeta(\lambda)+F_\zeta(-\lambda)\propto
 &\int_0^\infty dt~t^{-\zeta^2/2}\exp\left(-\frac{t}{8\alpha}-2\alpha \lambda^2
 t^{-\zeta}\right)\times\\
&\times~_1 F_1\left(1-\zeta/2,1/2,2\alpha \lambda^2
t^{-\zeta}\right)
\end{align}

Hence, the spectral density \eqref{rhoaftermanysubs} is determined
as the product of two integrals that are easily evaluated
numerically. The results for different values of $\zeta$ and
$\alpha=1$ are plotted in Fig. \ref{RhoNonGaussian}.

\end{document}